\documentclass[conference]{IEEEtran}
\usepackage{cite,graphicx,amsmath,amssymb}
\usepackage{subfigure}
\usepackage{citesort}
\usepackage{fancyhdr}
\usepackage{mdwmath}
\usepackage{mdwtab}
\usepackage{balance}
\usepackage{xcolor}
\usepackage{bm}

\usepackage{algorithm}
\usepackage{algorithmic}
\usepackage{multirow}
\usepackage{flafter}
\usepackage[
top    = 1.70cm,
bottom = 1.05in,
left   = 0.63 in,
right  = 0.63 in]{geometry}

\newtheorem{theorem}{Theorem}

\newtheorem{lemma}{Lemma}

\newtheorem{corollary}{Corollary}

\newtheorem{proposition}{Proposition}
\newtheorem{assumption}{Assumption}

\begin{document}
\title{Exploiting Multiple Access in Clustered Millimeter Wave Networks: NOMA or OMA?}

\author{
\IEEEauthorblockN{Wenqiang~Yi, Yuanwei~Liu and Arumugam Nallanathan } \IEEEauthorblockA{Queen Mary University of London, London, UK } }
\maketitle

\begin{abstract}
  In this paper, we introduce a clustered millimeter wave network with non-orthogonal multiple access~(NOMA), where the base station (BS) is located at the center of each cluster and all users follow a Poisson Cluster Process. To provide a realistic directional beamforming, an actual antenna pattern is deployed at all BSs. We provide a nearest-random scheme, in which near user is the closest node to the corresponding BS and far user is selected at random, to appraise the coverage performance and universal throughput of our system. Novel closed-form expressions are derived under a loose network assumption. Moreover, we present several Monte Carlo simulations and numerical results, which show that: 1) NOMA outperforms orthogonal multiple access regarding the system rate; 2) the coverage probability is proportional to the number of possible NOMA users and a negative relationship with the variance of intra-cluster receivers; and 3) an optimal number of the antenna elements is existed for maximizing the system throughput.
\end{abstract}
\vspace{-0.2cm}
\section{Introduction}
\vspace{-0.1cm}
The ever-increasing requirements of Internet-enabled applications and services have exhaustively strained the capacity of conventional cellular networks. One promising technology for augmenting the throughput of fifth generation (5G) wireless systems is exploiting new spectrum resources, e.g. millimeter wave~(mmWave)~\cite{rappaport2014millimeter}. Comparing with the traditional sub-6 GHz carrier frequency, mmWave has two distinguishing properties. One is the sensitivity to blockage effects, which increases the path loss of non-line-of-sight~(NLOS) signals~\cite{8016632}. The other is the small wavelength, which shortens the size of antenna elements so that large antenna arrays can be employed at mmWave devices for enhancing the directional array gain~\cite{Bai2015TWC}. Accordingly, several works have paid attention to these distinctive features when analyzing mmWave networks. The primary article~\cite{Bai2015TWC} proposed an line-of-sight~(LOS) disc model to reflect the impact of blockages. However, the directional beamforming method in this work was over-simplified and hence failed to depict the exact properties of a practical antenna. Then, a realistic antenna pattern was introduced in~\cite{7913628}. To capture the randomness of networks, stochastic geometry has been widely applied in numerous researches~\cite{Bai2015TWC,8016632,7593259,7913628,6489099}. More specifically, the locations of base stations~(BSs) follow a Poisson Point Process~(PPP). Since mmWave is able to support ultra-high throughput in short-distance communications~\cite{park2007short}, a recent work~\cite{8016632} considered a Poisson Cluster Process~(PCP) instead of PPP model to evaluate the short-range mmWave network, which obtains a close characterization to the real world.

In addition to expanding the available spectrum range, another significant objective of 5G cellular networks is improving the spectral efficiency. Recently, non-orthogonal multiple access~(NOMA) has kindled the attention of academia since it realizes multiple access in the power domain rather than the traditional frequency domain~\cite{7273963,8114722,7982794}. The main advantage of such approach is that NOMA possesses a perfect balance between coverage fairness and universal throughput~\cite{Zhiguo2015Mag}. In contrast to the conventional orthogonal multiple access~(OMA), a new technique named successive interference cancellation~(SIC) is applied at the NOMA user who has a robust channel condition to extract the requested information~\cite{7273963}. The power allocation strategies for NOMA networks were introduced in~\cite{7069272} to assure the fairness for all users. In a single cell scenario, the downlink sum-rate and outage probability were analyzed in~\cite{6868214} and the uplink NOMA performance with a power back-off method was investigated in~\cite{7390209}. However, the aforementioned articles focus on the noise-limited system and inter-cell interference is ignored for tractability of the analysis. In fact, the interference is an important factor when studying the coverage performance, especially in the sub-6~GHz networks. Authors in~\cite{7972929} offered a dense NOMA network with multiple inter-cells. Under this model, both uplink and downlink transmissions were evaluated. Like mmWave communications, stochastic geometry has also been utilized in NOMA networks~\cite{7812773,7445146} to model the positions of the primary and secondary NOMA users.

As mentioned above, although traditional NOMA networks can improve the spectral efficiency, the limited bandwidth resources below 6~GHz is substantially restrict the development of future wireless networks. Note that mmWave obtains a large amount of free spectrum. Applying NOMA into mmWave networks is an ideal way to satisfy the challenging demand of 5G cellular networks. Moveover, when the transmission distance is long, the path loss of mmWave communications is severer than that of conversional networks, so the inter-cell interference in mmWave networks is weak compared with sub-6~GHz scenarios'. Particularly,  in a loose mmWave network, such interference can be ignored due to its negligible received power, which will dramatically perfect the system coverage. With the aid of the PCP model as discussed in~\cite{8016632}, we proposed a practical user selection scheme to evaluate the performance of clustered mmWave networks with NOMA. The actual antenna pattern~\cite{7913628} is also employed to enhance the analytical accuracy. The main contributions of this treatise are as follows: 1)~we derive the closed-form equation for Laplace transform of inter-cluster interferences; 2)~novel expressions for coverage probabilities of near user and far user are deduced and closed-form algorithms under a loose network assumption are provided as well; 3)~NOMA performs better than OMA regarding the system rate in our mmWave networks due to the adjustable power allocation coefficients; and 4)~there exists an optimal number of antenna elements for obtaining the maximum system throughput due to the inverse feedback of two paired users.
\vspace{-0.3cm}
\section{Network Model}
\vspace{-0.2cm}
\subsection{Spatial Model}
\vspace{-0.2cm}
\begin{figure} [t!]
\centering
\includegraphics[width= 3.5in, height=1.85in]{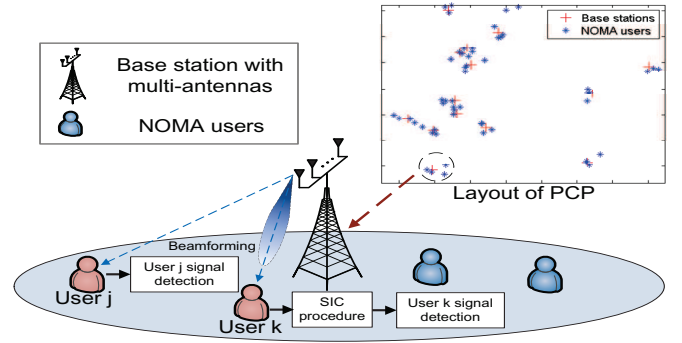}
\caption{Illustration of the clustered NOMA networks with mmWave communications. The spatial distributions of the NOMA users follow the PCP.}
\label{system_model}
\end{figure}
As shown in Fig.~\ref{system_model}, we consider a clustered downlink mmWave NOMA transmission scenario. All BSs are deployed at cluster centers and the distribution of NOMA users follows one of typical PCPs, which is a tractable variant of Thomas cluster process~\cite{7446343,8016632}. More specifically, the BSs constituting \emph{parent points} are distributed as a homogeneous PPP ${\Phi}$ with density ${\lambda_c}$. The union of NOMA receivers in a cluster with one parent at $y\in \Phi$ represents \emph{offspring points}, which is denoted by $\mathbb{N}_y$. Such NOMA users are independent and identically distributed (i.i.d.) following symmetric normal distributions with variance $\sigma^2$ and mean $0$. Therefore, the user at $x\in \mathbb{N}_y$ in reference to the corresponding BS at $y$ is given by
\begin{align}\label{1}
{f_X}\left( x \right) = \frac{1}{{2\pi {\sigma ^2}}}\exp \left( { - \frac{{{{\left\| x \right\|}^2}}}{{2{\sigma ^2}}}} \right).
\end{align}

In each cluster, the number of NOMA users is fixed as $2K$. These $2K$ users are divided into $K$ orthogonal pairs and each pair utilizes a single orthogonal beam. All BSs are assumed to serve the whole intra-cluster users at a same time. As a result, there is no intra-cluster interferences except the paired user's signal, but inter-cluster interferences from other BSs using the equivalent orthogonal beam are still existed. Additionally, a \emph{typical cluster} is randomly chosen to enhance the generality. The \emph{typical BS} included in the typical cluster is located at the origin of the considered plane. In this paper, we focus on a typical pair of users from the typical cluster, where the paired \emph{User~$k$} and \emph{User $j$} represent the near user and far user, respectively. A tractable user selection scheme named {Nearest-Random~(NR)} is introduced for analyzing the performance. More specifically, under NR scheme, User $k$ is the nearest receiver to the typical BS and User $j$ is randomly chosen from the rest of the NOMA users.
\vspace{-0.1cm}
\subsection{Blockage Effects}
\vspace{-0.1cm}
One remarkable characteristic of mmWave networks is that mmWave signals are sensitive to obstacles. Therefore, the LOS links have a distinctive path loss law with NLOS transmissions. Note that each cluster can be visualized as a dense mmWave network due to the small variance $\sigma^2$ of NOMA users. Under this condition, one obstacle may block all receivers behind it, so we adopt the LOS disc to model the blockage effect~\cite{Bai2015TWC,Andrews2015JSAC}. This blockage model fits the practical scenarios better than other patterns~\cite{7593259}. Accordingly, the LOS probability inside the LOS disc with a radius $R_L$ is one, while the NLOS probability outside the disc is one. With the aid of LOS disc model, we provide the path loss law of our proposed networks with a distance $\dot{r}$ as below
\begin{align}\label{2}
L_p(\dot{r}) = \mathbf{U}\left( {{R_L} - \dot{r}} \right){C_L}{\dot{r}^{ - {\alpha _L}}} + \mathbf{U}\left( {\dot{r }- {R_L}} \right){C_N}{\dot{r}^{ - {\alpha _N}}},
\end{align}
where $C_\kappa$ is the intercept and $\alpha_\kappa$ is the path loss exponent. $\kappa=L$ and $N$ represent the LOS and NLOS links, respectively. $\mathbf{U}(.)$ is the unit step function.
\vspace{-0.1cm}
\subsection{Antenna Beamforming}
\vspace{-0.2cm}
Another constraint for mmWave networks is the high cost and power consumption for signal processing components. We adopt analog beamforming in this work for achieving a low complexity beamforming design. More specifically, the directions of beams are controlled by phase shifters. We invoke the optimal analog precoding which implies that BSs try to align the direction of beams with the angel-of-departure (AoD) of channels. Hence high beamforming gains can be obtained. Assuming uniform linear array composed of $M$ antenna elements is deployed at all BSs, then based on this precoding design, the effective channel gain at User $k$ aligning with the optimal analog beamforming is given by
\begin{align}\label{7}
{\left| {{\bf{h}}_k^H{{\bf{w}}_k}} \right|^2} = {M}{ {{\left| {{g_k}} \right|}^2}},
\end{align}
where ${\bf{h}}_k$ and ${\bf{w}}_k$ are the channel vector and beamforming vector, respectively. $\left|g_k \right|$ represents the independent Nakagami-$N_\kappa$ fading for User $k$~\cite{Bai2015TWC}.

Regarding any other User $\hat k$, the effective channel gain is expressed as below
\begin{align}\label{8}
{\left| {{\bf{h}}_{\hat k}^H{{\bf{w}}_k}} \right|^2}=& \frac{{{{\left| {{g_{\hat k}}} \right|}^2}{{\sin }^2}\left( {\pi M\left( {{\theta _k} - {\theta _{\hat k}}} \right)} \right)}}{{M{{\sin }^2}\left( {\pi \left( {{\theta _k} - {\theta _{\hat k}}} \right)} \right)}} \nonumber \\
=& M{\left| {{g_{\hat k}}} \right|^2}{G_F}\left( {{\theta _k} - {\theta _{\hat k}}} \right).
\end{align}
where ${G_F}\left(  \cdot  \right)$ denotes the normalized \emph{Fej\'{e}r kernel} with parameter $\frac{1}{M}$ and $\left( {{\theta _k} - {\theta _{\hat k}}} \right)$ is uniformly distributed over $[-\frac{q}{\lambda},\frac{q}{\lambda}]$~\cite{7913628}.

\subsection{Signal Model}
We assume that in the typical cluster, the typical BS is located at $y_0\in \Phi$. Then, User $k$ at $x_k$ and User $j$ at $x_j$ are paired and served by the same beam. The distances of them obey $ {d_k} \le {d_j}$. Moreover, the power allocation coefficients satisfy the conditions that $a_k < a_j$ and $a_k + a_j = 1$, which is for fairness considerations~\cite{Zhiguo2015Mag}. In terms of other clusters, the interfering BS located at $y \in \Phi/y_0$ provides an optimal analog beamforming for User $\xi_y$, which is chosen uniformly at random. As a consequence, the received signal is given by
\begin{align}\label{signal user k}
{y_k} =& \underbrace {{\bf{h}}_k^H{{\bf{w}}_k}\sqrt {{a_k}{P_t}L_p\left( {\left\| {{x_k}} \right\|} \right)} {s_k}}_{{\bf{Desired}}\;{\bf{Signal}}} + \underbrace {{\bf{h}}_k^H{{\bf{w}}_k}\sqrt {{a_j}{P_t}L_p\left( {\left\| {{x_k}} \right\|} \right)} {s_j}}_{SIC\;{\bf{Signal}}} \nonumber\\
& + \underbrace {\sum\nolimits_{y \in \Phi /{y_0}} {{\bf{h}}_{y \to k}^H{{\bf{w}}_{{\xi _y}}}\sqrt {{P_t}L_p\left( {\left\| {{x_k} + y} \right\|} \right)} {s_{{\xi _y}}}} }_{{\bf{Inter}} - {\bf{Cluster}}} + \underbrace {{{\bf{n}}_0}}_{{\bf{Noise}}}
\end{align}
and
\begin{align}\label{signal user j}
{y_j} =& \underbrace {{\bf{h}}_j^H{{\bf{w}}_k}\sqrt {{a_j}{P_t}L_p\left( {\left\| {{x_j}} \right\|} \right)} {s_j}}_{{\bf{Desired}}\;{\bf{Signal}}} + \underbrace {{\bf{h}}_j^H{{\bf{w}}_k}\sqrt {{a_k}{P_t}L_p\left( {\left\| {{x_j}} \right\|} \right)} {s_k}}_{{\bf{Intra}} - {\bf{Cluster}}\;}\nonumber \\
  & + \underbrace {\sum\nolimits_{y \in \Phi /{y_0}} {{\bf{h}}_{y \to j}^H{{\bf{w}}_{{\xi _y}}}\sqrt {{P_t}L_p\left( {\left\| {{x_j} + y} \right\|} \right)} {s_{{\xi _y}}}} }_{{\bf{Inter}} - {\bf{Cluster}}} + \underbrace {{{\bf{n}}_0}}_{{\bf{Noise}}},
\end{align}
where $P_t$ is the transmit power of BSs. ${\bf{h}}_{y \to \varpi}$ represents the channel vector from BS at $y$ to User $\varpi$ and $\varpi \in \{k,j\}$.

Note that SIC is carried out at User $k$, and hence User $k$ will first decode the signal of User~$j$ with the following signal-to-interference-plus-noise-radio (SINR)
\begin{align}\label{SINR user k j}
{\gamma _{k \to j}} = \frac{{{a_j}{{\left| {{\bf{h}}_k^H{{\bf{w}}_k}} \right|}^2}L_p\left( {\left\| {{x_k}} \right\|} \right)}}{{{a_k}{{\left| {{\bf{h}}_k^H{{\bf{w}}_k}} \right|}^2}L_p\left( {\left\| {{x_k}} \right\|} \right) + {I_{{\rm{inter}},k}} + \sigma _n^2}},
\end{align}
where ${I_{{\rm{inter}},\varpi }} = \sum\nolimits_{y \in \Phi /{y_0}} {{{\left| {{\bf{h}}_{y \to \varpi }^H{{\bf{w}}_{{\xi _y}}}} \right|}^2}} {L_p}\left( {\left\| {{x_\varpi } + y} \right\|} \right)$ and $\sigma _n^2$ is the noise power normalized by $P_t$. Therefore, $1/\sigma _n^2$ is the signal-to-noise-ratio~(SNR) of our system.

If this decoding is successful, User $k$ then decodes the signal of itself. Based on \eqref{signal user k}, the SINR of  User $k$ to decode its own message can be expressed as
\begin{align}\label{SINR user k}
{\gamma _k} = \frac{{{a_k}{{\left| {{\bf{h}}_k^H{{\bf{w}}_k}} \right|}^2}L_p\left( {\left\| {{x_k}} \right\|} \right)}}{{{I_{{\rm{inter}},k}} + \sigma _n^2}}.
\end{align}

Regarding User $j$, it directly decodes its own message by treating the signal of User $k$ as the interference. Based on \eqref{signal user j}, the SINR of  User $j$ is given by
\begin{align}\label{SINR user j}
{\gamma _j} = \frac{{{a_j}{{\left| {{\bf{h}}_j^H{{\bf{w}}_k}} \right|}^2}L_p\left( {\left\| {{x_j}} \right\|} \right)}}{{{a_k}{{\left| {{\bf{h}}_j^H{{\bf{w}}_k}} \right|}^2}L_p\left( {\left\| {{x_j}} \right\|} \right) + {I_{{\rm{inter}},j}} + \sigma _n^2}}.
\end{align}
\vspace{-0.3 cm}
\section{Distance Distributions}
\vspace{-0.1 cm}
In this section, we will discuss the distance distributions of the typical paired users. These distributions are the basis for analyzing the performance of our system. To simplify the notation, we first introduce a typical distribution named \emph{Rayleigh Distribution} as below~\cite{7446343}.

\emph{Rayleigh Distribution}: Under Rayleigh distribution, the probability density function~(PDF) is given by ${R_p}\left( v, \sigma \right) = \frac{v}{{{\sigma ^2}}}\exp \left( { - \frac{v^2}{{2{\sigma ^2}}}} \right),v > 0$ and the cumulative distribution function is given by ${R_c}\left( {v,\sigma } \right) = 1 - \exp \left( { - \frac{{{v^2}}}{{2{\sigma ^2}}}} \right),v > 0$, where $\sigma^2$ is the variance parameter as mentioned in~\eqref{1}.
\vspace{-0.1 cm}
\subsection{Distance Distribution of Near User}
\vspace{-0.1 cm}
In the typical cluster, we assume that the distances between NOMA users and the typical BS form a set $\{R_i\}_{i=1:2K}$ which can be denoted by $\mathbb{R}_{y_0}$. The realization of $R_i$ is defined as ${r_i} = \left\| {x_i} \right\|$, where $x_i \in \mathbb{N}_{y_0}$. Note that $x_i$ is i.i.d. as a Gaussian random variable with $\sigma^2$. If the considered NOMA user is selected at random, we are able to drop the index $i$ from $r_i$ as every $r_i$ follows the same distribution. Under this condition, $r$ is a Gaussian random variable with variance $\sigma^2$, so the PDF of distance $r$ is as follows~\cite{8016632}
\begin{align}
f_r\left(r\right)={R_p}\left( v, \sigma \right).
\end{align}

Under NR scheme, User $k$ is the nearest NOMA receiver. We assume User $k$ is located at $x_1$ with a distance $r_1$ to the typical BS, so the PDF of $r_1$ can be expressed as below.
\begin{lemma}\label{lemma1}
In NF scheme, the near user is the nearest NOMA user with a distance $r_1=\left\|x_1\right\|$. Therefore, the PDF of such distance $r_1$ is as follows
\begin{align}\label{nearest}
f_d^1\left( {{r_1}} \right) = \frac{{2K{r_1}}}{{{\sigma ^2}}}\exp \left( { - \frac{{Kr_1^2}}{{{\sigma ^2}}}} \right).
\end{align}
\begin{IEEEproof}
\emph{Note that there are $(2K-1)$ NOMA users distributed farther than the considered distance $r_1$. Therefore, the PDF of distance $r_1$ is $f_d^1\left( {{r_1}} \right) = 2K{\left( {1 - {R_c}\left( {{r_1},\sigma } \right)} \right)^{2K - 1}}{R_p}\left( {{r_1},\sigma } \right)$. With the aid of Rayleigh distribution, we are able to derive the expression as above.}
\end{IEEEproof}
\end{lemma}
\vspace{-0.1 cm}
\subsection{Distance Distribution of Far User}
\vspace{-0.1 cm}
In contrast to the near user, the far user in NR scheme is randomly chosen from the rest NOMA users in the typical cluster. Assuming the possible User $j$ is located at $x_f^r \in \mathbb{N}_{y_0} / x_1$ with a distance $r_f = \left\| {x_f^r} \right\|$, the distribution of distance $r_f$ is expressed as below.
\begin{lemma}\label{lemma2}
The randomly selected far user in NF scheme at $x_f^r$ has a distance $r_f$ to the typical BS and $r_f>r_1$, so the conditional PDF of distance $r_f$ is given by
\begin{align}
f_d^f\left( {{r_f}}{|{r_1}} \right) = \left\{ {\begin{array}{*{20}{c}}
   {\frac{{{R_p}\left( {{r_f},\sigma } \right)}}{{1 - {R_c}\left( {{r_1},\sigma } \right)}}}, & {{r_f} > {r_1}}  \\
   0, & {{r_f} \le {r_1}}  \\
\end{array}} \right..
\end{align}
\begin{IEEEproof}
\emph{When ${{r_f} \le {r_1}}$, the probability is zero as all far users are farther than $r_1$. When $r_f>r_1$, the possible User $j$ follows Rayleigh distribution over the rang $(r_1,\infty ]$. Therefore, such distance distribution can be summarized as above.}
\end{IEEEproof}
\end{lemma}
\vspace{-0.1 cm}
\section{Performance Evaluation}
\vspace{-0.1 cm}
In this section, we characterize the coverage performance and system throughput of proposed NR strategy depending on the distributions of distances.

In NR scheme, we choose the nearest NOMA receiver as near user due to the best channel condition, which contributes to minimizing the coefficient $a_k$ so that more power can be allocated to far user. On the other side, selecting User $j$ at random from the rest NOMA users aims to provide a fair selection law. To make the tractable analysis, we first deduce the \emph{Laplace Transform of Interferences} in our system.
\vspace{-0.1 cm}
\subsection{Laplace Transform of Interferences}
\vspace{-0.1 cm}
We only concentrate on the Laplace transform of inter-cluster interferences because there is no interfering devices located in the typical cluster.
\begin{lemma}\label{lemma5}
The inter-cluster interferences are provided from all BSs except the typical BS, then the closed-form Laplace transform of such interferences is given by
\begin{align}
{\mathcal{L}_{I}}\left( s \right)  = \exp \left( { - \frac{{{\pi ^2}{\lambda _c}R_L^2}}{{{2n_1}}}\sum\limits_{{i_1} = 1}^{{n_1}} {\mathcal{G}_F^I\left( s,{\frac{{{\zeta _{{i_1}}}q}}{\lambda }} \right)\sqrt {1 - \zeta _{{i_1}}^2} } } \right),
\end{align}
where
\begin{align}
& \mathcal{G}_F^I\left( s,g \right) = {\rho _N}\left( {\frac{{sM{C_N}{G_F}\left( g \right)}}{{{N_N}R_L^{{\alpha _N}}}}} \right) - {\rho _L}\left( {\frac{{{N_L}R_L^{{\alpha _L}}}}{{sM{G_F}\left( g \right){C_L}}}} \right), \\
&{\rho _L}\left( v \right) = \frac{{_2{F_1}\left( {{N_L},{N_L} + \frac{2}{{{\alpha _L}}};{N_L} + \frac{2}{{{\alpha _L}}} + 1; - v} \right)2{v^{{N_L}}}}}{{\left( {{\alpha _L}{N_L} + 2} \right)}},\\
&{\rho _N}\left( v \right) = {}_2{F_1}\left( { - \frac{2}{{{\alpha _N}}},{N_N};1 - \frac{2}{{{\alpha _N}}}; - v} \right),(\alpha_N>2),
\end{align}
${}_2F_1(.)$ is Gauss hypergeometric function. ${\zeta _{{i_1}}} = \cos \left( {\frac{{2{i_1} - 1}}{{2{n_1}}}\pi } \right)$ over $[-1,1]$ denotes the Gauss-Chebyshev node and $i_1=1,2,...,n_1$. The parameter $n_1$ has a function to balance the complexity and accuracy~\cite{7445146}. Only if the $n_1\rightarrow \infty$, the equality is established.
\begin{IEEEproof}
\emph{See Appendix A.}
\end{IEEEproof}
\end{lemma}
\vspace{-0.1 cm}
\subsection{Coverage Probability for Near User}
\vspace{-0.1 cm}
We pre-decide SINR thresholds $\tau_k$ and $\tau_j$ for User $k$ and User $j$, respectively. These two thresholds should satisfy the condition $({a_j} - {\tau _j}{a_k} > 0)$ to ensure the success of NOMA protocols~\cite{7445146}. Note that the near user has SIC procedure, which means the decoding will be success only when $({\gamma _{k \to j}}>\tau_j)$. Under this condition, the coverage probability is the percentage of the received SINR $\gamma_k$ that excesses $\tau_k$. Accompanying with fact that the near user under NR scheme is the nearest NOMA receiver, the coverage probability for near user User $k$ can be defined as below
\begin{align}
P_k\left( {{\tau _k}} \right) = P\left[ {{\gamma _k} > {\tau _k}|{\gamma _{k \to j}} > {\tau _j},{r_1} = \left\| {{x_1}} \right\|} \right],
\end{align}
where $\mathbb{P}[.]$ is the probability function.

The expressions for coverage probability of near user can be divided into two case : ${a_k}{\tau _j}<{a_j} \leq {a_k}{\tau _j}( {1 + \frac{1}{{{\tau _k}}}} )$ and ${a_j} > {a_k}{\tau _j}( {1 + \frac{1}{{{\tau _k}}}} )$. In most mmWave scenarios, the power allocation coefficient $a_j$ is far larger than $a_k$ because of the severe path loss for far user. Therefore, we focus on the more practical case with ${a_j} > {a_k}{\tau _j}( {1 + \frac{1}{{{\tau _k}}}} )$ in this paper.
\begin{theorem}\label{theorem1}
When ${a_j} > {a_k}{\tau _j}( {1 + \frac{1}{{{\tau _k}}}} )$, the coverage probability of near user under NR scheme is given by
\begin{align}\label{31}
P_k\left( {{\tau _k}} \right) \approx & \int_0^{{R_L}} {{\Theta _L}\left( {{r_1},{\tau _k},{a_k}} \right)f_d^1\left( {{r_1}} \right)d{r_1}}  \nonumber \\
&+ \int_{{R_L}}^\infty  {{\Theta _N}\left( {{r_1},{\tau _k},{a_k}} \right)f_d^1\left( {{r_1}} \right)d{r_1}} .
\end{align}
where
\begin{align}
{\Theta _\kappa }\left( {r,\tau ,\beta } \right) =& {\sum\limits_{{n_\kappa } = 1}^{{N_\kappa }} {\left( { - 1} \right)} ^{{n_\kappa } + 1}}{N_\kappa \choose n_\kappa}\exp \left( { - \frac{{{n_\kappa }{\psi _\kappa }\tau {r^{{\alpha _\kappa }}}\sigma _n^2}}{{\beta M{C_\kappa }}}} \right)\nonumber\\
&\times {\mathcal{L}_I}\left( {\frac{{{n_\kappa }{\psi _\kappa }\tau {r^{{\alpha _\kappa }}}}}{{\beta M{C_\kappa }}}} \right),
\end{align}
and ${\psi _\kappa } = {N_\kappa }{\left( {{N_\kappa }!} \right)^{ - 1/{N_\kappa }}}$.
\begin{IEEEproof}
\emph{With the similar proof procedure of Theorem 1 in~\cite{Bai2015TWC}, we obtain the expression as above.}
\end{IEEEproof}
\end{theorem}

In the reality, the coverage radius of the macro BS is always farther than $R_L$, which means the majority of BSs communicates with the considered user via an NLOS link. Note that the received power from NLOS signals is negligible~\cite{Bai2015TWC} . We propose a special case as below.

\emph{Special Case 1:} In a loose network, the density of BSs is small enough to ensure the majority of BSs utilizing NLOS links to provide the inter-cluster interferences. Therefore, we ignore all inter-cluster interferences and coverage probability from NLOS links, namely, ${\mathcal{L}_I}(.)=0$ and $\Theta _N(.)=0$. Moreover, we assume $\alpha_L=2$ as it is the practical value for numerous frequencies~\cite{deng201528}.

\begin{corollary}\label{corollary3}
Under special case 1, the closed-form coverage probability for near user is changed into
\begin{align}
\hat P_k\left( {{\tau _k}} \right) \approx \frac{K}{{{\sigma ^2}}}{\sum\limits_{{n_L} = 1}^{{N_L}} {\left( { - 1} \right)} ^{{n_L}}}{N_L \choose n_L}\frac{{\exp \left( { - A\left( {{\tau _k}} \right)R_L^2} \right) - 1}}{{A\left( {{\tau _k}} \right)}},
\end{align}
where $A\left( \tau  \right) = \frac{{{n_L}{\psi _L}\tau \sigma _n^2}}{{{a_k}M{C_L}}} + \frac{K}{{{\sigma ^2}}}$.
\begin{IEEEproof}
\emph{With the fact $\int_0^B {v\exp \left( { - A{v^2}} \right)dv}  = \frac{{1 - \exp \left( { - A{B^2}} \right)}}{{2A}}$, \eqref{31} can be simplified as above.}
\end{IEEEproof}
\end{corollary}
\subsubsection{Coverage Probability for Far User}
In contrast to the near user, the coverage probability for User $j$ at $x_j$ only depends on $\tau_j$. However, as the directional beamforming of the typical BS is aligned towards the near user, the effective channel gain for User $j$ will fit \eqref{8} rather than \eqref{7}. Note that the far user is randomly selected from the intra-cluster NOMA receivers, except the near user. We define the coverage probability of far user as follows
\begin{align}
P_j\left( {{\tau _j}} \right) = \mathbb{P}\left[ {{\gamma _j} > {\tau _j}\left| {{r_f} = \left\| {{x_j}} \right\|} \right.} \right].
\end{align}
As discussed in \emph{Lemma~\ref{lemma2}} and Laplace transform of interferences, we obtain the coverage probability of User~$j$ as below.
\begin{theorem}\label{theorem2}
The coverage probability for User $j$ at $x_j$ with a distance $r_f$ is given by
\begin{align}
P_j\left( {{\tau _j}} \right) \approx \frac{\pi }{{2{n_2}}}\sum\limits_{{i_2} = 1}^{{n_2}} {\mathcal{G}_j\left( {{\tau _j},\frac{{{\zeta _{{i_2}}}q}}{\lambda }} \right)} \sqrt {1 - \zeta _{{i_2}}^2},
\end{align}
where
\begin{align}
{\cal G}_j\left( {{\tau _j},g} \right) \approx & \int_0^{{R_L}} {\int_{{r_1}}^{{R_L}} {{\Delta _L}\left( {{\tau _j},g} \right)} d{r_f}} f_d^1(r_1)d{r_1} \nonumber \\
&+ \int_{{R_L}}^\infty  {\int_{{r_1}}^\infty  {{\Delta _N}\left( {{\tau _j},g} \right)} d{r_f}}f_d^1(r_1) d{r_1}, \\
{\Delta _\kappa }\left( {{\tau _j},g} \right) = &{\Theta _\kappa }\left( {{r_f},{\tau _j},\left( {{a_j} - {\tau _j}{a_k}} \right){G_F}\left( g \right)} \right)f_d^f\left( {{r_f}} |r_1\right).
\end{align}
\begin{IEEEproof}
\emph{With the aid of Lemma 2, we are able to obtain the above expressions via the similar method as discussed in Theorem 1.}
\end{IEEEproof}
\end{theorem}
\begin{corollary}\label{corollary5}
Under special case 1, in a loose network, the closed-form coverage probability for far user is given by
\begin{align}
\hat P_j\left( {{\tau _j}} \right) \approx \frac{\pi }{{2{n_2}}}\sum\limits_{{i_2} = 1}^{{n_2}} {\hat {\mathcal{G}}_j\left( {{\tau _j},\frac{{{\zeta _{{i_2}}}q}}{\lambda }} \right)} \sqrt {1 - \zeta _{{i_2}}^2},
\end{align}
where
\begin{align}
&\hat {\mathcal{G}}_j\left( {{\tau _j},g} \right) = \sum\limits_{{n_L} = 1}^{{N_L}} {{{\left( { - 1} \right)}^{{n_L} + 1}}} {N_L \choose n_L}\frac{K}{{2{\sigma ^4}Q\left( {{\tau _j},g} \right)}}\nonumber \\
&\times \Bigg( {\frac{1}{{Q\left( {{\tau _j},g} \right) + \chi }} + \frac{{Q\left( {{\tau _j},g} \right)\exp \left( { - \left( {Q\left( {{\tau _j},g} \right) + \chi } \right)R_L^2} \right)}}{{\left( {Q\left( {{\tau _j},g} \right) + \chi } \right)\chi }} }\nonumber \\
& - {\exp \left( { - Q\left( {{\tau _j},g} \right)R_L^2} \right)} \Bigg),
\end{align}
and $Q\left( {{\tau _j},g} \right) = \frac{{{n_L}{\psi _L}{\tau _j}\sigma _n^2}}{{\left( {{a_j} - {\tau _j}{a_k}} \right){G_F}\left( g \right)M{C_L}}} + \frac{1}{{2{\sigma ^2}}}$ and $\chi  = \frac{{2K - 1}}{{2{\sigma ^2}}}$.
\begin{IEEEproof}
\emph{With the similar proof method as discussed in {Corollary~\ref{corollary3}}, we obtain {Corollary~\ref{corollary5}}.}
\end{IEEEproof}
\end{corollary}
\vspace{-0.1 cm}
\subsection{System Throughput}
\vspace{-0.1 cm}
To compare with the traditional OMA method, we provide the system throughput in this part. Assuming the bandwidth $B$ is separated equally into two parts for transferring information to User $k$ and User $j$ under OMA. We have the system throughput expressions for NOMA and OMA as below.
\begin{proposition}
If the rate requirement for User $k$ and User $j$ are $R_k$ and $R_j$, respectively, the equations of system throughput for NOMA and OMA are given by
\begin{align}
&R_s^{NOMA} = {R_k}{P_k}( {1 - {2^{\frac{{{R_k}}}{B}}}} ) + {R_j}{P_j}( {1 - {2^{\frac{{{R_j}}}{B}}}} ), \\
 &R_s^{OMA} = {R_k}{P_k}( {1 - {2^{\frac{{{2R_k}}}{B}}}} )| {_{{a_k} = 1}} + {R_j}{P_j}( {1 - {2^{\frac{{{2R_j}}}{B}}}} )| {_{{a_j} = 1}}.
\end{align}
\end{proposition}
\vspace{-0.1 cm}
\section{Numerical Results}
\vspace{-0.1 cm}
In this section, we first present the general network settings in Table.~\ref{table1}~\cite{Bai2015TWC,8016632}. The reference distance is assumed to be one meter. Then, numerical results with Monte Carlo simulations are provided for analyzing the performance of our networks.

\begin{figure}
  \centering
  \includegraphics[width= 3.2 in, height=1.85 in]{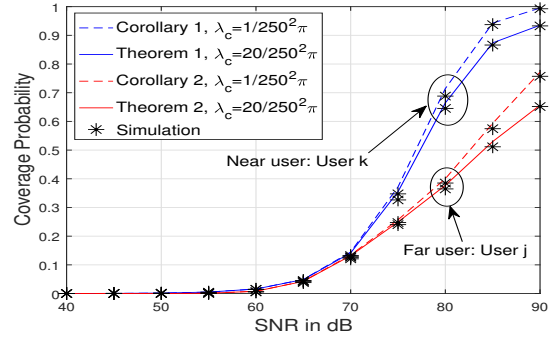}\\
  \caption{Coverage probabilities for the typical paired users versus SNR, with different densities of BSs $\lambda_c$.}\label{fig1}
\end{figure}

As shown in Fig.~\ref{fig1}, there is a negligible difference between the theoretical results and the simulations, thereby corroborating our analysis. In the loose network, namely $\lambda_c=1/250^2\pi$, closed-form expressions in \emph{Corollary 1} and \emph{Corollary 2} can be the replacement of the exact analytical equations due to the easy-operation and high-accuracy. Moreover, these closed-form expressions are suitable for numerous practical scenarios, where the density of macro BSs is around $1/250^2\pi$. On the other side, the coverage probabilities for two typical paired users are proportional to SNR and near user outperforms far user under our assumptions.
\begin{figure}
  \centering
  \includegraphics[width= 3.2 in, height=1.85 in]{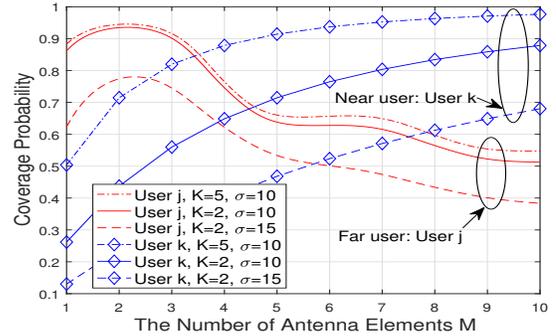}\\
  \caption{Coverage probabilities for the typical paired users versus the number of antenna elements, with different $K$, $\sigma$, $\lambda_c=1/250^2\pi$ and SNR$=83$~dB.}\label{fig2}
\end{figure}
\begin{figure}
  \centering
  \includegraphics[width= 3.2 in, height=1.85 in]{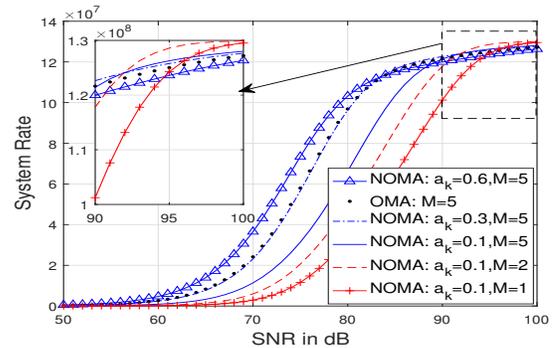}\\
  \caption{System rate in bps for NOMA and OMA versus SNR, with different $a_k$, $M$, $R_k=10^8$ bps, $R_j=3 \times 10^7$ bps and $\lambda_c=1/250^2\pi$. }\label{fig3}
\end{figure}
\begin{table}[ht]
\centering
\caption{General Network Settings}
\label{table1}
\begin{tabular}{|l|l|}
\hline
   LOS disc range     & $R_L=100$ m\\ \hline
   Density of BSs     & $\lambda_c=1/ (250^2\pi)$ m$^{-2}$\\ \hline
   Path loss law for LOS     & $\alpha_L=2$, $N_L=3$\\ \hline
   Path loss law for NLOS     & $\alpha_N=4$, $N_N=2$\\ \hline
   Number of antennas     & $M=10$\\ \hline
   Carrier frequency     &  $f_{m}=28$ GHz\\ \hline
   Power allocation coefficients & $a_k=0.1$, $a_j=0.9$ \\ \hline
   Variance                  & $\sigma^2=100 $ \\ \hline
   Number of NOMA users in a cluster  & $2K=4$ \\ \hline
   SINR thresholds    & $\tau_k=1$, $\tau_j=0.2$ \\ \hline
   Bandwidth per resource block & $B=100$ MHz \\ \hline
   Antenna parameter  & $\lambda=4q$ \\ \hline
\end{tabular}
\end{table}

In terms of the antenna beamforming, two paired users have inverse performances as illustrated in Fig.~\ref{fig2}. In general, the coverage probability of near user is a increasing function with the antenna scale $M$, while that of far user is opposite. Due to the randomness of the beamforming vector ${\bf{w}}_j$, the coverage probability of User $j$ is fluctuant. In our case, $M=2$ becomes the best choice for far user. Lastly, the large number of NOMA users $K$ and small variance $\sigma^2$ are able to improve the coverage performance for both users.

Comparing with the OMA method, NOMA scenario with $a_k=0.6$ performs better in the low SNR region, while in the high SNR region, the best option is NOMA method with $a_k=0.1$. Accordingly, by modifying the power allocation coefficients in NOMA protocol, we are able to achieve a higher system rate than utilizing the traditional OMA technique. Fig.~\ref{fig3} also demonstrates that the impact of antenna scale is various across the considered SNR range. More elements deployed at the antenna will weaken the throughput at the high SNR region. As a result, there should be an optimal $M$ for maximizing the system rate.
\vspace{-0.1 cm}
\section{Conclusion}
\vspace{-0.1 cm}
In this paper, we propose a NR scheme in clustered mmWave networks with NOMA technique. With the aid of the stochastic geometry, the novel analytical expressions for coverage and system throughput are presented, especially we derive closed-form equations for a loose network, which can be utilized in numerous practical noise-limited scenarios. As analyzed in previous sections, the coverage probability is proportional to SNR and the number of NOMA users $K$. Large variance $\sigma^2$ will impair the received SINR. Lastly, our NOMA system beats the traditional OMA case regarding the system rate by adjusting the power allocation coefficients.
\vspace{-0.1 cm}
\numberwithin{equation}{section}
\section*{Appendix~A: Proof of Lemma~\ref{lemma5}} \label{appendix_A}
\renewcommand{\theequation}{A.\arabic{equation}}
\setcounter{equation}{0}
\vspace{-0.1 cm}
For User $k$, the Laplace transform of interferences is
\begin{align}\label{A.1}
{\mathcal{L}_I^k}\left( s \right) =& \mathbb{E}\left[ {\exp \left( { - s{I_{{{inter},k}}}} \right)} \right]\nonumber \\
 =& \mathbb{E}\left[ {{e^{ - s\sum\limits_{y \in \Phi /{y_0}} {M{{\left| {{g_{y \to k}}} \right|}^2}{G_F}\left( {{\theta _k} - {\theta _{{\xi _y}}}} \right){L_p}\left( {\left\| {{x_k} + y} \right\|} \right)} }}} \right]\nonumber \\
 =&\mathcal{L}_L^k(s)\mathcal{L}_N^k(s).
\end{align}

By applying the probability generating functional of PPP~\cite{stoyanstochastic} and calculating the expectation of Gamma random variable $\left| {{g_{y \to k}}} \right|^2$ and the antenna beamforming $G_F(.)$, we obtain
\begin{align}\label{A.2}
{\cal L}_L^k\left( s \right) = {e^{ - \frac{{\pi {\lambda _c}\lambda }}{q}\int_{ - \frac{q}{\lambda }}^{\frac{q}{\lambda }} {\left( {\int_0^{{R_L}} {\left( {1 - {{\left( {1 + \frac{{sM{G_F}\left( g \right){C_L}}}{{{N_L}{v^{{\alpha _L}}}}}} \right)}^{ - {N_L}}}} \right)vdv} } \right)} dg}}.
\end{align}

Using the same proof procedure for NLOS group, we obtain
\begin{align}\label{A.3}
{\cal L}_N^k\left( s \right) = {e^{ - \frac{{\pi {\lambda _c}\lambda }}{q}\int_{ - \frac{q}{\lambda }}^{\frac{q}{\lambda }} {\left( {\int_{{R_L}}^\infty  {\left( {1 - {{\left( {1 + \frac{{sM{G_F}\left( g \right){C_N}}}{{{N_N}{v^{{\alpha _N}}}}}} \right)}^{ - {N_N}}}} \right)vdv} } \right)} dg}}.
\end{align}

By substituting \eqref{A.2}, \eqref{A.3} and the definition of Gauss hypergeometric function into \eqref{A.1}, we obtain
\begin{align}\label{A.4}
\mathcal{L}_I^k\left( s \right) = \exp \left( { - \frac{{\pi {\lambda _c}\lambda R_L^2}}{2q}\int_{ - \frac{q}{\lambda }}^{\frac{q}{\lambda }} {\mathcal{G}_F^I\left( s,g \right)} dg} \right).
\end{align}

Note that the Laplace transform of interferences for User~$j$ has the similar deducing procedure, so two paired users share the same expressions. Therefore we are able to drop the index $k$ from \eqref{A.4}. After that, applying Gaussian-Chebyshev quadrature equation into \eqref{A.4}, the proof is complete.
\vspace{-0.1 cm}
\bibliographystyle{IEEEtran}

\end{document}